\documentclass[12pt]{article}

\usepackage{amsmath}
\usepackage{url}
\usepackage{graphics}
\usepackage{graphicx,color}
\usepackage[authoryear]{natbib}
\usepackage{amsfonts, animate,subfigure}
\usepackage{amssymb,dsfont}
\usepackage[margin=1in]{geometry}
\usepackage{blkarray}
\usepackage{setspace}
\usepackage[colorlinks, urlcolor = blue, citecolor = blue]{hyperref}
\usepackage[ruled]{algorithm2e}
\usepackage{bm}
\usepackage[format=hang,labelfont=bf,font={small,it}]{caption}
\usepackage[capposition=below]{floatrow}
\usepackage{paralist}
\usepackage{soul}

\usepackage{color}

\usepackage[rightbars, color]{changebar}
\cbcolor{blue}
\begin{document}

\title{Inferring social structure from continuous-time interaction data}

\author{Wesley Lee\footnote{Contact email: wtlee@uw.edu}\\{\small University of Washington} \and Bailey K. Fosdick\\{\small Colorado State University} \and Tyler H. McCormick\\{\small University of Washington}}
\date{}
\maketitle

\begin{abstract}

Relational event data, which consist of events involving pairs of actors over time, are now commonly available at the finest of temporal resolutions. Existing continuous-time methods for modeling such data are based on point processes and directly model interaction ``contagion,'' whereby one interaction increases the propensity of future interactions among actors, often as dictated by some latent variable structure. In this article, we present an alternative approach to using temporal-relational point process models for continuous-time event data.  We characterize interactions  between a pair of actors as either spurious or as resulting from an underlying, persistent connection in a latent social network. We argue that consistent deviations from  expected  behavior, rather than solely high frequency counts, are crucial for identifying well-established underlying social relationships.  This study aims to explore these latent network structures in two contexts: one comprising of college students and another involving barn swallows.\\


\end{abstract}

\doublespacing

\section{Introduction} \label{Introduction}

Data consisting of interactions between pairs of actors (i.e. dyads) over time are commonly referred to as relational event data \citep{butts2008relational} and are being collected at unprecedented rates.  As mobile technology becomes nearly ubiquitous, it is increasingly possible to measure social interactions with arbitrarily granular temporal precision.  Call detail records (CDRs), for example, provide detailed descriptions of mobile phone interactions on a national scale \citep{blumenstock2012inferring}. The widespread adoption of social media has led to a wealth of continuous-time activity data   \citep{sadilek2012modeling}. In addition, in ecology, advances in tracking devices have resulted in an increase in animal telemetry logs documenting  movement and interactions between animals \citep{rutz2009new}.

In business, the Enron email corpus, which consists of hundreds of thousands of emails sent between employees over a multiyear period, has been used to study organizational communication and its adaptation and evolution during times of crises \citep{diesner2005communication}. On a coarser temporal scale, informal transactions between households in developing countries are hypothesized to allow these agents to mitigate the risks of potential economic shocks \citep{fafchamps2003risksharing}. In related literature, interbank loans can serve as insurance against region-specific liquidity shocks \citep{allen2000financial}.

Relational event data are distinct from traditional, survey-based network measures in two key ways. First, relational events do not directly quantify relationships between actors, but serve as indicators of these relations and embed underlying social structure and dynamics. Second, relational event data are typically obtained from administrative logs or other sources not explicitly designed for research.  This feature presents additional statistical challenges, but also means that collecting relational event data is often relatively inexpensive compared to collecting survey data and can allow researchers to study larger populations of actors at higher temporal resolutions and for more extended periods \citep{watts2007twenty}.

Analyses of relational event data typically involve representing interactions in data structures congruent with existing discrete-time dynamic social network models \citep{sarkar2005dynamic, durante2014nonparametric, krivitsky2014separable, sewell2015latent}.  To mirror the form of traditional survey network data, interactions are commonly aggregated into a series of weighted adjacency matrices, also called sociomatrices, whose elements represent the number of interactions between each pair of actors within fixed time intervals. Taken together, the sequence of adjacency matrices can be viewed as a weighted network which evolves in discrete-time. Inference on  discrete-time dynamic networks is typically performed using models that characterize the manner in which dyadic interactions change from one time interval to the next, often assuming a Markov process whereby the network at time $t$ only depends on its past states through the network at time $t-1$ \citep{snijders2005models, sewell2016latent}. 

One issue with the aforementioned aggregation approach is the implicit assumption that unobserved meaningful social (network) relations are easily ascertained from the observed noisy interaction data. Continuous-time interaction data consists only of measurements \emph{when individuals interact}, and the absence of interaction should not be taken as explicit declaration of no relationship. More generally, interaction counts between dyads should not necessarily be taken as a direct measure of the strength of the underlying dyad relationship. Consider email records in which employee A emailed coworker B multiple times per week and emailed his/her manager C once every other week. From our perspective, both of these email patterns may indicate strong relations between employee A and the others. Although communication with the manager is relatively infrequent compared to the communication with the coworker, the relationship between A and his/her manager should be classified as strong and significant.

\begin{figure}[ht]
\centering
\includegraphics[scale=0.7]{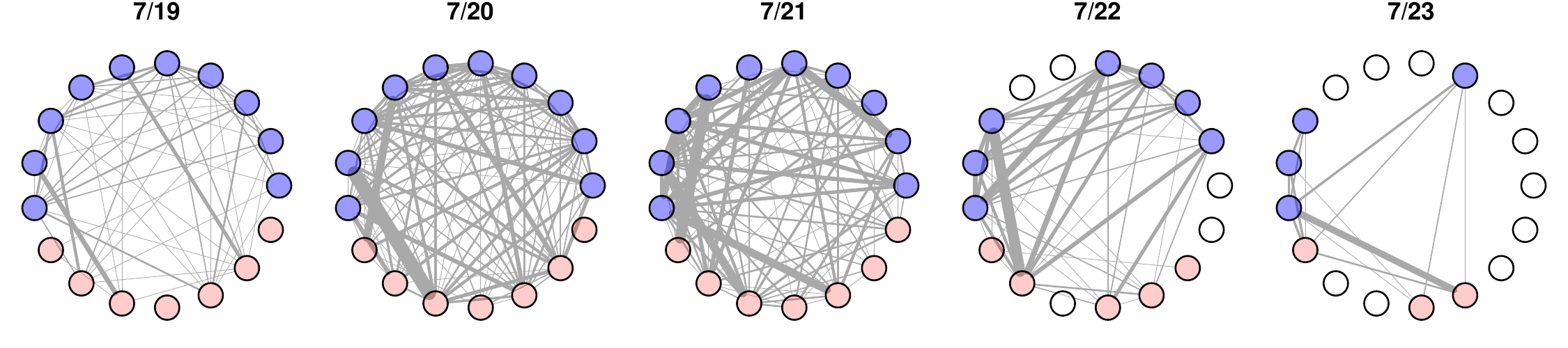} 
\caption{\small Daily snapshots of aggregated barn swallow interactions. Nodes are colored by sex (females red; males blue) and edges widths are proportional to the number of interactions between each pair.  Birds deemed to be unmonitored on a given day (see Appendix~\ref{Swallow Appendix: Data}) are shown in white.}
\label{fig:snap}
\end{figure}

Another key drawback of performing  discrete-time data aggregation when modeling continuous-time interaction data is that the time intervals are often chosen arbitrarily and conclusions can be greatly impacted by these choices \citep{sulo2010meaningful, Sekara23082016}. In addition, the temporal dynamics of the interactions are possibly lost in the aggregation process depending on the length of the time interval. Consider Figure~\ref{fig:snap}, for example, where proximity data from a barn swallow study has been aggregated over each day (see Sections~\ref{Intro: Applications} and \ref{Birds: Data} for further details on this data).
By aggregating to the level of the day we preclude the possibility of capturing intricate within-day social dynamics. While smaller time intervals are able to capture more of these dynamics, models with Markov structure may be inappropriate for modeling the resulting series of weighted networks.  Conditional dependence among behavior across time would then be more likely to span multiple intervals. Returning to the example of call records, an assumption of Markovian dynamics for summaries of behavior at, for example, an hourly level would be unsuitable since associated individuals are unlikely to call each other on such a frequent basis.

In this article, we propose a continuous-time approach to modeling relational event data that explicitly separates interactions from underlying social relations. Specifically, we conceptualize continuous-time interaction data as representing \emph{manifestations of latent network structure}. The model we propose possesses two distinctive properties. First, rather than viewing the data as direct observations of the network of interest, we assume observed interactions arise from a point processes with propensity influenced by the latent network structure. It is the dynamics of this \emph{underlying} social network that we argue is typically most informative about population social structure and of direct interest to researchers. Second, we avoid decisions on temporal resolution by modeling both the observed interactions and dynamics of the latent network structure in continuous-time. The statistical challenge, therefore, is to infer the underlying network structure and its evolution through time. In this paper we assume interactions and connection in the latent network are undirected, although our methods could readily be extended to handle directed interactions and networks.

Existing work in continuous-time modeling for relational data focuses on \emph{excitatory} point processes \citep{simma2012modeling, blundell2012modelling, perry2013point}. Hawkes processes~\citep{hawkes1971spectra} are one such process; in a Hawkes process, the occurrence of an event causes an immediate increase in intensity for that process and/or other Hawkes processes. These processes are favored for their ability to model reciprocity and transitivity in relational event data. Despite being used to model continuous-time data, existing point process models assume a static underlying network \citep{blundell2012modelling, linderman2014discovering} or eschew modeling a network altogether, focusing on just modeling the interaction data \citep{dubois2010modeling, simma2012modeling, perry2013point}.

In this article, our primary focus is on inferring an underlying dynamic network, as opposed to inferring casual dependencies between  the noisy interactions. In this sense, our work parallels that of \citep{linderman2014discovering} and \citep{scharf2015dynamic}, which  aim to infer network relations from individual-level movement and behavior.  We posit that the patterns in relational events are primarily governed by temporal and covariate-dependent factors and only secondarily by excitatory triggers. Thus, we model relational events as arising from inhomogeneous Poisson processes with intensities dependent on time, covariates, and an underlying dynamic latent network. In the interest of parsimony, if two individuals lack a relationship in the  underlying network, we consider their interactions as the result of spurious behavior and thus use a relatively simple model.  Model complexity is instead focused on interactions between pairs of individuals inferred to have a  connection in the underlying network.   

The remainder of the paper is organized as follows.  In Section~\ref{Intro: Applications}, we briefly introduce two applications for our relational event framework involving face-to-face interactions among college students and animal telemetry data. In Section~\ref{Model}, we describe the general framework in detail, and in Section~\ref{Inference} we propose a Bayesian inference procedure for models from this framework. We apply our framework to the two aforementioned datasets in Sections~\ref{Students} and \ref{Birds}.  Codes to replicate the analyses and figures presented in this paper are available at \url{https://wesleytlee.github.io/relational-event-networks/}.

\subsection{Proximity interactions: students and swallows} \label{Intro: Applications}

Using our method, we explore network structure in two settings with data collected using proximity sensors: one involving interactions between college students and another consisting of contacts among barn swallows (\textit{Hirundo rustica erythrogaster}).

First, we consider proximity information for 57 undergraduate students living in a dormitory at MIT during the 2008-2009 school year \citep{madan2012sensing}. Participants were given Bluetooth-enabled smartphones, and time stamps of Bluetooth (proximity) pings between nearby phones were recorded. In addition, students were surveyed five times throughout the school year and asked about assorted measures of their health and relationships with other subjects in the study. This dataset has been used to study the spread of infection \citep{dong2012graph} and the proliferation of health-related choices among these undergraduates \citep{madan2010social}. In both studies, researchers found physical contact information derived from the proximity data played a critical explanatory role beyond that of the self-reported relationships. These findings imply that information obtained from human behavior can inherently differ from that in reported relationships, and the relative importance and validity of each information type may vary across applications.

Next, we examine interactions within a population of Colorado barn swallows collected over three days during mating season \citep{levin2015performance, levin2016stress}.  Barn swallows are small birds often found nesting in man-made structures.  Due to their small size and speed in flight, recording proximity-based social interactions by visual observation is infeasible. Proximity loggers have emerged as promising alternative tools for collecting animal interaction data in species of all sizes. The deployment of these tools presents new opportunities to gain understanding of animal social structure.  We investigate a set of close encounters between barn swallows outfitted with proximity loggers.

\section{A Continuous-Time Interaction Framework} \label{Model}

We propose a hierarchical modeling framework for continuous-time relational event data in which dyad interactions are modeled using an inhomogeneous Poisson process with intensity dependent on dyadic covariates and the state of the dyad in an underlying latent network. The latent network relation for each dyad is binary and modeled with a slow-changing continuous-time Markov chain. This construction allows our framework to express two properties of primary interest: 1) We avoid directly equating interaction frequency with relational strength in the network by detaching the process of modeling the relational events from the process of modeling the social network.  Consequently, actors that are connected in the latent network can assume interaction patterns with varying frequencies. 2) Both the interactions and the network are modeled directly in continuous-time so that no choices about temporal resolution of data aggregation are required.  Furthermore, we retain both the flexibility to capture fine temporal dynamics and the computational benefits of modeling the network evolution as Markovian.

For every pair of actors $(i,j)$, denote the vector of interaction event times as $\bm{t}^{ij}\equiv\left\{t_{1}^{ij},...,t_{N_{ij}}^{ij}\right\} $ and the time interval on which these events have the potential to occur by $\bm{T}^{ij}\equiv\left[T_{1}^{ij},T_{2}^{ij}\right)$. Additionally, denote the set of dyadic, possibly time-varying, covariates as $x_{ij}(t)$. For the MIT data,  covariates of potential interest include whether students reside on the same floor of the dormitory or are in the same year of school. For the swallow data, covariates include sex pairings and similarities between phenotypic traits. 

We model the vector of interaction times $\bm{t}^{ij}$ for dyad $(i,j)$ using an inhomogeneous Poisson process with intensity $\lambda_{ij}(t)$, representing the instantaneous rate of an interaction occurring between actors $i$ and $j$ at time $t$.  The log probability of events occurring at times $\bm{t}^{ij}$ for dyad $(i,j)$ is then given by
\begin{equation}
	\log p
	\Big(\bm{t}^{ij}\big|\lambda_{ij}(t)\Big)
	=\sum_{t\in\bm{t}^{ij}}\log\lambda_{ij}(t) 
	- \int_{\bm{T}^{ij}}\lambda_{ij}(t)dt.
	\label{eq:Poisson log-likelihood}
\end{equation}
Conditional on the underlying intensity function, the presence/absence of interactions in disjoint time intervals are independent. Additionally, we assume  interactions for different dyads arise independently conditional on their intensity functions. In contrast to Hawkes processes, dependence between dyad interactions is therefore entirely captured by the dyad intensities.  Using $y_{ij}(t)$ as an indicator of whether a network connection exists between $i$ and $j$ at time $t$, we model each intensity $\lambda_{ij}(t)$ as dependent on the time $t$, dyad-specific covariates $x_{ij}(t)$, and the latent connection $y_{ij}(t)$.  

Introducing time and covariate dependencies of the event data into the Poisson intensity function modulates the influence of these factors in the estimated network and allows the model to differentially weigh interactions as evidence of network structure. Interactions occurring at specific times and/or under certain circumstances may be much more probable in the presence of a network connection rather than in the absence thereof, and should be accounted for accordingly. We structure $\lambda_{ij}(t)$ as follows:
\begin{equation}
	\lambda_{ij}(t)=\Bigg(1+m\Big(t,x_{ij}(t)\Big)y_{ij}(t)\Bigg)\times
	\Bigg[w\Big(t,x_{ij}(t)\Big)+a\Big(t,x_{ij}(t)\Big)y_{ij}(t)\Bigg],
	\label{eq:Poisson Process Intensity}
\end{equation}
and denote the corresponding sets of model parameters for functions $\{m, w, a\}$ as $\left\{c_{m},c_{w},c_{a}\right\}$. Assuming no connection in the latent network, $y_{ij}(t)=0$, the intensity function reduces to a baseline intensity $w(t,x_{ij}(t))$, dependent on time and dyad-specific covariates. The function $w$ corresponds to the rate of ``spurious" interactions, which result when actors interact due to circumstance. For example, students in the MIT dormitory may be in close proximity to one another at night due solely to room assignments. We leave the exact form of $w$, as well as the other functions, to be specified on an case-by-case basis using domain expertise and descriptive summaries of the data.

Behavior for connected dyads can differ from the baseline in two ways: they may generally tend to interact more often, through the multiplicative factor $m$, and their interactions may have a different pattern, through an additive adjustment to the baseline $a$. When inferring the latent network, interactions in line with expected behavior for connected dyads but not expected behavior for unconnected dyads are suggestive of a latent connection. Thus, both evidence of generally higher activity levels than baseline and interactions that deviate from the baseline pattern according to our adjustment $a$ suggest an underlying relationship.

Poisson processes model interactions as instantaneous events in time. However, for many potential applications, including both the MIT and barn swallow proximity data, encounters can last for non-negligible durations. In these cases, we remove the time while the event is taking place from the time interval $\bm{T}^{ij}$ for each given pair, preventing the model from modeling the probability of another event while an interaction is taking place and assuring that the intensity is calibrated to the appropriate time window.  Future work could extend our modeling framework to use the duration length as a potential source of information about future interactions and/or network relations.

The latent network $\bm{Y} \equiv \left\{y_{ij}(t)\,: \, i,j \in \{1,...,n\}, \, i\not=j, \,t\in \bm{T}^{ij}\right\}$ can be decomposed into dyad-specific paths $\left\{y_{ij}(t)\,:\,t\in \bm{T}^{ij}\right\}$ representing the state of a connection between dyad $(i,j)$ over time. These paths $\{y_{ij}(t)\}$ are modeled as independent stationary continuous-time Markov chains (CTMCs) taking values in $\{0,1\}$. Each path $y_{ij}\left(t\right)$ is characterized by a sparsity parameter $s > 0$ and an transition parameter $q > 0$. The probability of transitioning from one network state to another in any interval of length $t$ is given by the entries of the following transition matrix:
\begin{equation}
{\LARGE P} = \;\; 
\begin{blockarray}{ccc}
 &  0 &  1\\
\begin{block}{c[cc]}
0 & \left(1-s\right)+se^{-qt} & s-se^{-qt}\\
1 & \left(1-s\right)-\left(1-s\right)e^{-qt} & s+\left(1-s\right)e^{-qt} \\
\end{block}
\end{blockarray}.
\label{eq:CTMC Transition Matrix}
\end{equation}
This Markov chain has stationary distribution equal to $(1-s,s)$, meaning that, asymptotically, $y_{ij}(t)=1$ proportion $s$ of the time. Thus, the parameter $s$ governs the overall propensity for a connection between any dyad, while $q$ governs the speed at which the network changes, i.e. relations form/dissolve. Modeling each binary relation as a CTMC with small $q$ values yields a slow-changing network. 

In addition, we allow the CTMC parameters to depend on non-time-varying dyadic covariates, i.e. $s=s(x_{ij})$ and $q=q(x_{ij})$ with respective sets of parameters $\left\{c_{s},c_{q}\right\}$. Allowing for the sparsity parameter to depend on dyadic covariates allows for the introduction of some network structure, such as block models. For example, one could cluster individuals in the inferred network into known communities (such as floors in the MIT data) by allowing the sparsity parameter to positively depend on whether or not individuals in each dyad shared the same community.

The hierarchical nature of the model allows us to separate the process of modeling a network from the process of modeling interactions. The interaction model acts as a low-pass filter on the data, distinguishing between spurious interactions and those suggestive of a meaningful relationship. Evidence of network relations between actors is more generally interpreted through deviance from the expected baseline level of activity rather than only through high raw interaction counts. The hidden CTMC under the interaction model smooths evidence across time so that relationships are thus characterized by consistent, prolonged deviations from baseline behavior.

\section{Bayesian Inference} \label{Inference}

We propose a Bayesian sampling procedure for estimating the parameters of the model $\bm{\theta}\equiv\left\{c_{m},c_{w},c_{a},c_{s},c_{q}\right\}$ and the latent network $\bm{Y}= \left\{y_{ij}(t)\,: \, i,j \in \{1,...,n\}, \, i\not=j,\,t\in \bm{T}^{ij}\right\}$.  After appropriately eliciting priors, we are interested in the posterior distribution of the parameters and network conditional on the data $D\equiv\left\{\bm{t}^{ij}: \, i,j \in \{1,...,n\}\right\}$. Although the posterior distribution is analytically intractable, we can approximate it to an arbitrary degree of accuracy by taking samples from the posterior with a Markov chain Monte Carlo (MCMC) algorithm.  MCMC algorithms involve constructing a Markov chain with stationary distribution equal to the posterior distribution of the parameters and network given the data.

Inference proceeds by first sampling the parameters and then estimating the network. The posterior distribution of the parameters $\bm{\theta}$ given the data $D$ is defined
\begin{equation}
	p\left(\bm{\theta}|D\right)\propto
	\left[ \prod_{\left(i,j\right)}
	p\left(\bm{t}^{ij}|\bm{\theta}\right) \right]
	\times p\left(\bm{\theta}\right).
	\label{eq:Posterior for Parameters}
\end{equation}
After obtaining a sample of size $K$ from this posterior, the posterior probability that dyad  $\left(i,j\right)$ is connected in the network at time $t$ can then be approximated as 
\begin{align}
	p\big(y_{ij}(t)=1|D\big) &\approx
	\frac{1}{K}\sum_{k=1}^{K}p\big(y_{ij}(t)=1|D,\bm{\theta}^{\left(k\right)}\big),
	\label{eq:MC Network Approx}
\end{align}
where $\bm{\theta}^{\left(k\right)}$ is the $k$th sample of $\bm{\theta}$ from the posterior. The main challenge in this procedure is calculating $p\left(\bm{t}^{ij}|\bm{\theta}\right)$, the probability of a dyad's interactions  conditional only on the parameters, since doing so requires marginalizing over the the dyad's network path $\left\{y_{ij}(t)\,:\,t\in \bm{T}^{ij}\right\}$.  This marginalization is intractable for our model due to the continuous-time nature of the Poisson process data and the infinite number of possible network paths. 

\begin{figure}[ht]
    \centering
    \includegraphics[scale=1]{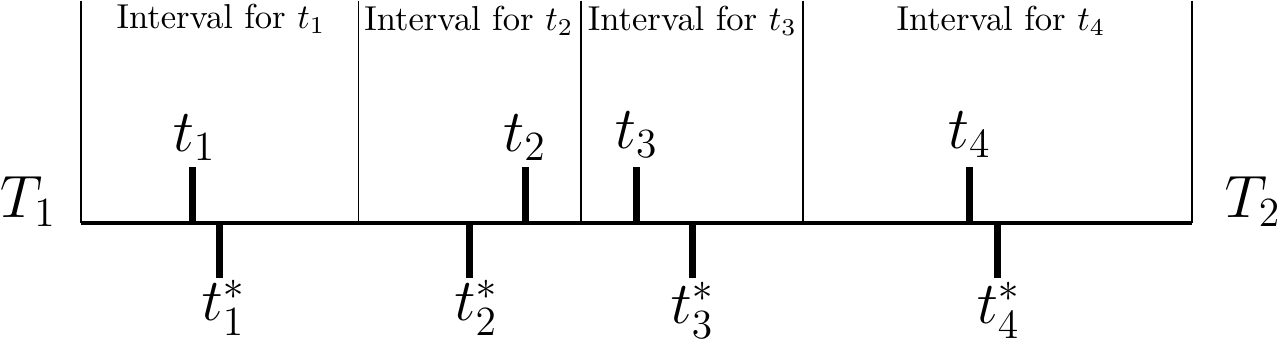}
    \caption{Sample partitioning of $(T_1,T_2)$ based on four interactions.  Above the horizontal time axis are the time points $\{t_1,t_2,t_3,t_4\}$ at which events were observed.  Below the axis are the midpoints of each interval $\{t_1^*,t_2^*,t_3^*,t_4^*\}$.  In the interest of parsimony, the network is estimated at each of the midpoint locations and assumed constant within each interval when calculating the probability of the observed data.}
    \label{fig:Binning}
\end{figure}

To circumvent this integration issue, we approximate the probability of each dyad's interactions given the network and parameters by discretizing the underlying network path and restricting potential transitions to a fixed number of times dependent on the observed interactions. Namely, we partition the full time interval $\bm{T}^{ij}$ based on the data, such that each (sub)interval consists of a single interaction and the surrounding times that are closer to this interaction than any other (see Figure \ref{fig:Binning}). We then restrict the network relation to be constant within each of these intervals for the purpose of calculating the probability of the observed interactions. The state of the network edge under this partition is characterized by Markov behavior from the midpoint of one subinterval to the next. The object of inference then changes from a continuous-path $\left\{ y_{ij}(t)\,:\,t\in\bm{T}^{ij}\right\}$ to the approximation $\left\{ y_{ij}(t_{l}^{ij,*})\right\} _{l=1}^{N_{ij}}$, where $t_{l}^{ij,*}$ is the midpoint of the interval containing $t_{l}^{ij}$. 
To infer the state of the path at other times $t$, we simply interpolate between the probabilities of a relation from the nearest two midpoints.  Since only one interaction is recorded within each interval and the network is assumed to be slow-changing, we do not expect the underlying network to change repeatedly between interval midpoints.

The procedure we use is not the only possible discretization procedure; for example, one might partition all time intervals $\bm{T}^{ij}$ into preset, equally spaced intervals. However, one advantage of our chosen method is that it automatically trades off between parsimony and flexibility based on the interaction behavior for a dyad.  Tying potential state changes for a network path to interactions focuses modeling effort on periods where there is the most information (arising through interactions) in the data and spends less effort modeling periods of low activity, where there is comparatively little information in the data.

We can view $\left\{y_{ij}\left(t_{l}^{ij,*}\right)\right\}_{l=1}^{N_{ij}}$ as a hidden Markov model (HMM) with ``emissions" $\bm{t}^{ij}$ arising from a Poisson process with intensity $\lambda_{ij}(t)$. That is, we see evidence of the evolution of the approximated network path, which itself is unobserved, through the relational event data observed in each interval, and borrow strength across intervals to ensure the path is appropriately slow-changing. The ``forward-backward algorithm" is a method for estimating the hidden states in an HMM \citep{rabiner1989tutorial}.  Using the forward variables from this algorithm, we can marginalize over the latent path to obtain an approximation to the probability $p\left(\bm{t}^{ij}|\bm{\theta}\right)$. We can then approximate the posterior $p\left(\bm{\theta}|D\right)$ in \eqref{eq:Posterior for Parameters}, and, in turn, obtain posterior samples $\left\{\bm{\theta}^{\left(k\right)}\right\}$ and subsequently estimate $\left\{y_{ij}\left(t\right)\,:\,t\in\bm{T}^{ij}\right\}$ at the midpoints of the intervals $\left\{ t_{l}^{ij,*}\right\} _{l=1}^{N_{ij}}$ using the forward-backward algorithm.

We use Stan \citep{carpenter2016stan} to sample from the posterior distribution of the parameters and latent network.  See the Appendices \ref{MIT Appendix: Model} and \ref{Swallow Appendix: Model} for more details on this procedure.

\section{Proximity Interactions among College Students} \label{Students}

In this section, we analyze interactions between college students from the MIT Social Evolution dataset \citep{madan2012sensing}, as mentioned in Section \ref{Intro: Applications}. We describe the MIT data in more detail and motivate the model selected. We illustrate one of the advantages of our inferred network over the network snapshots provided by the survey data.

\subsection{The Data} \label{Students: Data}

The data consists of Bluetooth proximity information collected from the phones of 57 MIT undergraduates living in a single dormitory over the period of a school year. Although political leanings and health measures were collected by the surveys, most basic personal information about the students is unavailable (e.g. sex), with the exception of what floor each student lived on and what year of study they were in. 

\begin{figure}[ht]
	\centering
	\includegraphics[scale=0.7]{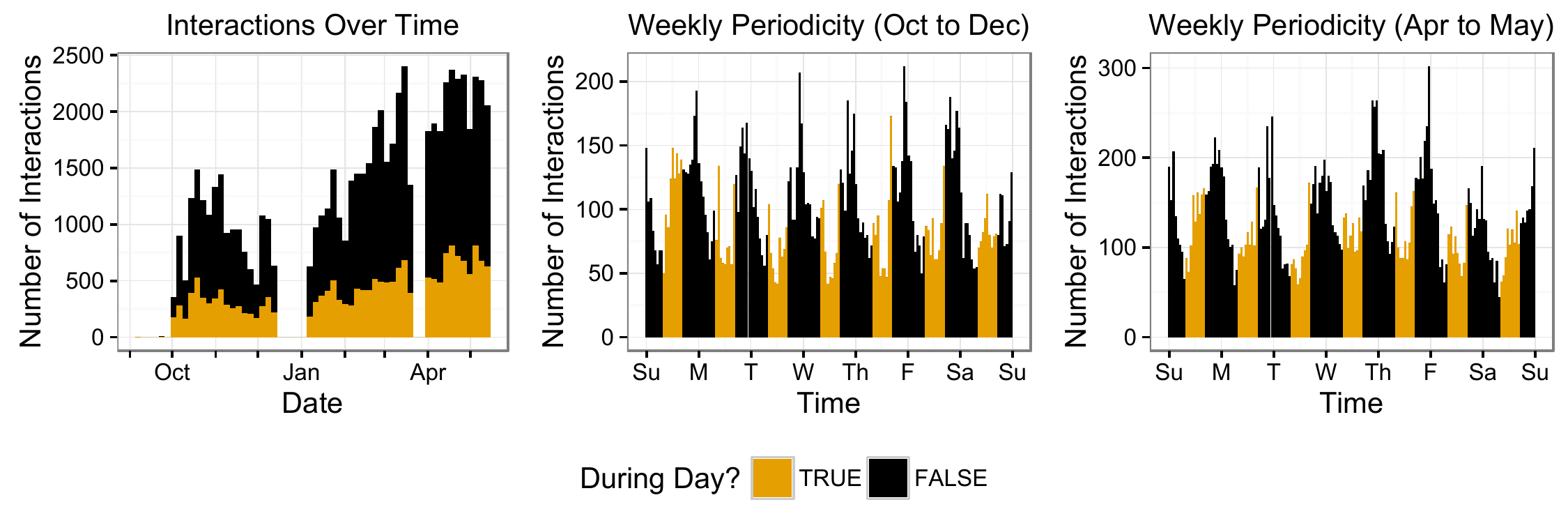} 
	\caption{Descriptive plots of the MIT Bluetooth proximity interactions. Left: 
		Distribution of interactions over the school year. Center: Interaction frequencies within a week, aggregated from October to December for each dyad. Right: Interaction frequencies within a week, aggregated from April to May.}
	\label{fig:Descriptive MIT}
\end{figure}

After appropriate data cleaning, as detailed in Appendix~\ref{MIT Appendix: Data}, the dataset consists of 66,432 Bluetooth-based proximity interactions among the 1,596 dyads. We model each dyad only over the interval when both students' phones are active (sending/receiving Bluetooth signals) in the proximity logs. Summaries of the data are plotted in Figure \ref{fig:Descriptive MIT} and reveal strong temporal dependencies. The left panel shows the rate of logged interactions increases throughout the school year. The center and right panels suggests strong daily and weekly periodicity: the majority of interactions occur at night, when students are more likely to be in their dormitory, and interactivity during the weekend appears to deviate from weekday behavior, with less of a disparity between daytime and nighttime activity levels.

\subsection{The Model} \label{Students: Model}

Figure \ref{fig:Descriptive MIT} reveals both weekly periodicity patterns and an increase in activity level as the year progresses. To model the former, we approximate the weekly periodicity with components of its Fourier series representation, choosing the number of sine waves to trade off between simplicity and flexibility. We approximate the trend of increased overall interactivity over the school year by splitting up the year into three terms: first semester (Oct. - Dec.), second semester before spring break (Jan. - March), and second semester after spring break (April - May),  and allowing for different activity levels in each period. The similarities between the center and right panels in Figure \ref{fig:Descriptive MIT} suggest that the pattern of weekly periodicity is largely unrelated to the increase in overall interactivity over the school year.

Dyadic covariates and friendship relations between students are associated with weekly patterns of interactivity and thus are important to explore further in order to build a reasonable model. One unique aspect of the MIT data is that we can leverage the survey data in model construction. At the level of the dyad, only a weak relation exists between the number of interactions recorded and how often friendship was reported in surveys (see Figure \ref{fig:MIT Ints vs Surveys} in Appendix \ref{MIT Appendix: Data}). However, when aggregating across dyads, we find significant differences in the interactivity patterns between reported friends and non-friends. Using survey data as a proxy for friendship, we compare the weekly patterns for reported friends/non-friends and individuals who live on the same/different floors, where we consider a dyad to be friends if at least one student named the other as a ``close friend".
\begin{figure}[ht]
	\centering
	\includegraphics[scale=0.7]{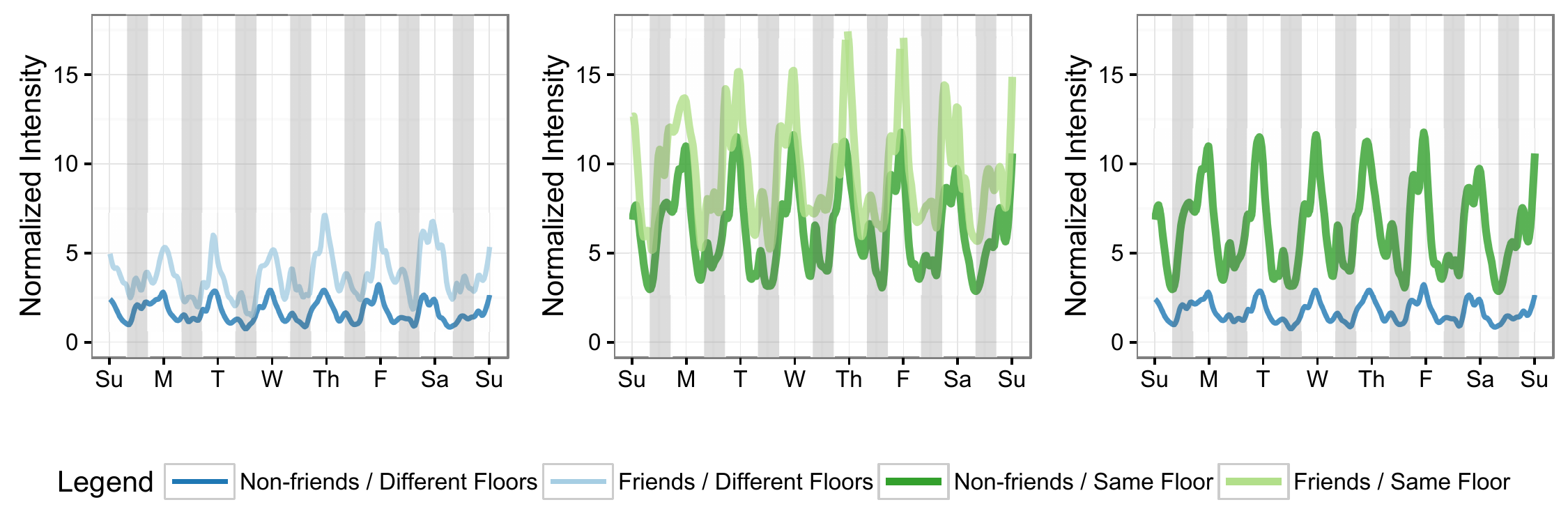} 
	\caption{A comparison of weekly interactivity patterns by survey-reported friendship and whether or not pairs of individuals reside on the same floor. Daytime is highlighted with grey bars. Intensities are normalized so that non-friends on different floors have mean intensity of one. In order to avoid potential friendship dynamics, we consider a pair of students ``friends" if at least one of the two identified themselves as ``close friends" in four of the five surveys conducted. We consider a pair of students ``non-friends" if they jointly identified themselves as ``close friends" once or less.}
	\label{fig:MIT Weekly Patterns}
\end{figure}
In the right panel of Figure \ref{fig:MIT Weekly Patterns}, non-friends that live on the same floor tend to interact with higher propensity than non-friends that on different floors. However, while their overall interactivity levels differ, individuals in these two groups interact in similar weekly patterns. The left and center panels suggest that friends interact more than non-friends regardless of floor status, although the relative compounding effect is stronger for friends who live on different floors. Additionally, we find some evidence that friends may interact more than non-friends during the daytime (defined here to be 8 AM to 5 PM). Since the students all live in the same dormitory, encounters during the day, when more conscious effort is required to be in close proximity to others, may be especially indicative of friendship.

Using the notation in \eqref{eq:Poisson Process Intensity}, we construct the following interaction model:
\begin{align}
	w\left(t,\text{floor}(i),\text{floor}(j)\right) & = c_{0,\text{term}}
	\left(1+c_{1}\mathbf{1}_{\text{floor}(i)=\text{floor}(j)}\right)
	\left(k_{0}+\sum_{l=1}^{3}k_{l,1}\cos\left(k_{l,2}t+k_{l,3}\right)\right)
	\label{eq:MIT Baseline}\\
	m\left(\text{floor}(i),\text{floor}(j)\right) & =c_{2,1}\mathbf{1}_{\text{floor}(i)\neq\text{floor}(j)}
	+c_{2,2}\mathbf{1}_{\text{floor}(i)=\text{floor}(j)}
	\label{eq:MIT Multiplier}\\
	a\left(t,\text{floor}(i),\text{floor}(j)\right) & = c_{0,\text{term}}
	\left(1+c_{1}\mathbf{1}_{\text{floor}(i)=\text{floor}(j)}\right)
	c_{3}\mathbf{1}_{\text{daytime}}\left(t\right)
	\label{eq:MIT Adjustment}
\end{align}
In the baseline $w$, three sets of sinusoidal terms model the weekly behavior while multipliers for school term and floor govern the overall levels of interactivity. We characterize friendship by an increase in daytime activity relative to the baseline pattern of weekly periodicity and an overarching multiplicative increase in intensity, dependent on floor status. Based on Figures \ref{fig:Descriptive MIT} and \ref{fig:MIT Weekly Patterns}, we assume this increased propensity to interact during the day for friends is subject to the same term- and floor-dependent modifiers that characterize general interactivity between dyads and use the same multipliers $c_{0,\text{term}}$ and $c_{1}$ in the daytime adjustment \eqref{eq:MIT Adjustment} that govern the overall interactivity in the baseline \eqref{eq:MIT Baseline}.

We can similarly use the survey data to inform our model for the network paths, treating the reported relationships as observations of CTMCs at the survey dates.
\begin{figure}[ht]
	\centering
	\includegraphics[scale=0.7]{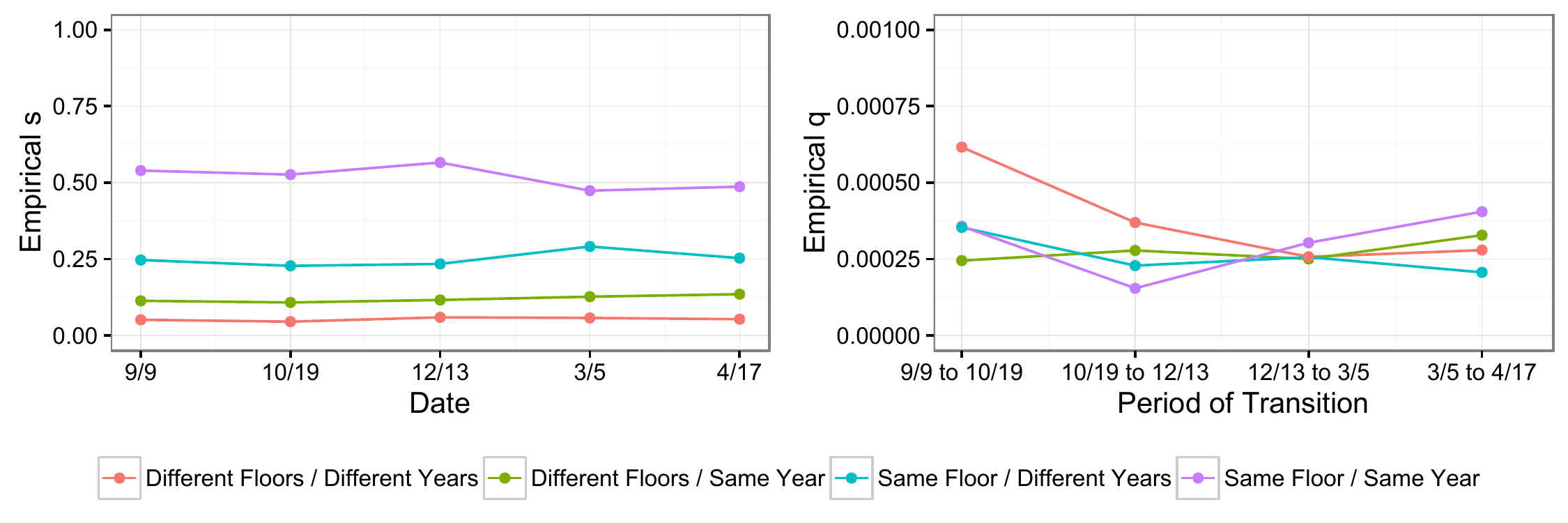} 
	\caption{Empirical CTMC parameter values for the four possible dyadic floor/year pairings derived from survey data. Sparsity $s$ is estimated by the proportion of dyads who report friendship at each date, while transition parameter $q$ is derived from the proportion of dyads which change reported friendship between consecutive survey dates.}
	\label{fig:MIT Survey CTMC}
\end{figure}
In Figure \ref{fig:MIT Survey CTMC}, living on the same floor and being in the same year have independent, multiplicative effects on the empirical probability of friendship for dyads. Additionally, these proportions are constant throughout the school year and changes between survey dates are consistent with a single, low transition rate $q$. We model the network paths as CTMCs with sparsity
\begin{equation}
	s\left(\text{floor}(i),\text{floor}(j),
	\text{year}(i),\text{year}(j)\right)=
	\left(1+s_{1}\mathbf{1}_{\text{floor}(i)=\text{floor}(j)}\right)
	\left(1+s_{2}\mathbf{1}_{\text{year}(i)=\text{year}(j)}\right)s_{0}
	\label{eq:MIT Friendship Sparsity}
\end{equation}
and constant transition rate $q$. Introducing this structure into the sparsity model encourages block model behavior in the underlying network, where students who live on the same floor or are in the same year are more likely to be friends with one another and hence form ``blocks" of friendship.

Before we can obtain estimates of the student network, we must assign priors to the parameters of our model. We observe a strong cyclical pattern in communications (associated with, for example, day and night), so we fix $k_{i,2}$ using a Fourier transform on the observed weekly pattern of interactivity. We set the interactivity multiplier for the first school term equal to one ($c_{0,\text{t1}}=1$) to ensure the identifiability of these multipliers. We specify weakly informative priors on the remaining parameters based on support matching; additional details on these prior distributions are presented in Appendix~\ref{MIT Appendix: Model}.

\subsection{Network Analysis} \label{Students: Analysis}

We used Stan to generate 9,000 samples (after burn-in) of the parameters from the posterior distribution. Details of the sampling procedure are provided in Appendix~\ref{MIT Appendix: Model}.

\begin{figure}[ht]
	\centering
	\includegraphics[scale=0.7]{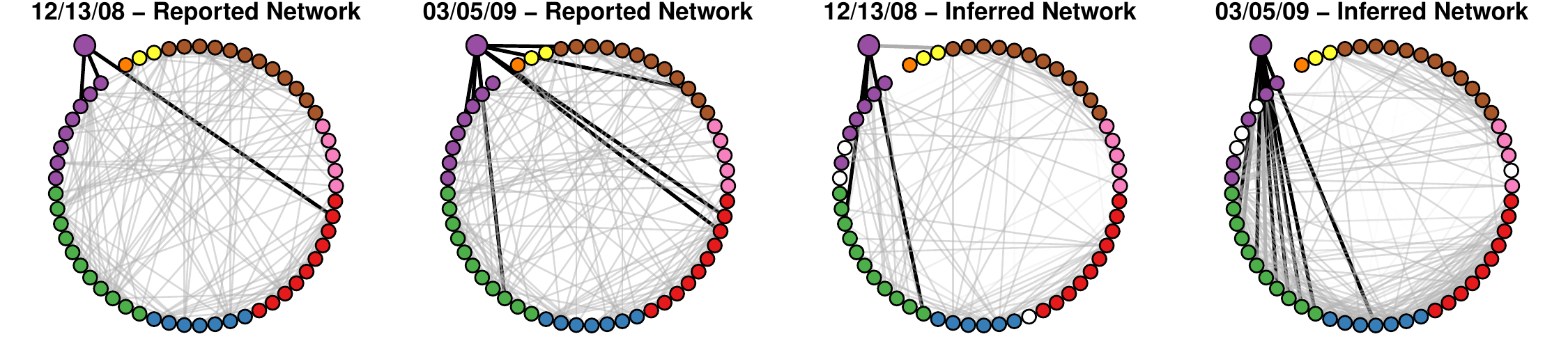} 
	\caption{Snapshots of networks of the MIT students at two survey dates. Students are colored by the floor they lived on. For the reported network, students are considered friends if either reports the other as a ``close friend." In the case of the inferred networks, students are colored white if their phones are inactive and we can no longer infer relations for them. Uncertainty in the inferred relations is quantified by the opacity of the ties. In all snapshots, student \#2 is highlighted, as well as his/her relations.}
	\label{fig:MIT Snapshots}
\end{figure}

One of the key benefits of the inferred network is that it evolves in continuous-time, allowing for a more detailed examination of network evolution. For example, in the left two panels of Figure \ref{fig:MIT Snapshots} we plot two snapshots of the MIT network as self-reported by students.
In the first survey date shown, December 2008, student \#2 (highlighted, towards the top left) was reported to be friends with three other individuals. By March of the following year, there were eight friendships involving this student. Between the two dates, six friendships were formed and one dissolved. Yet, with only this information, we have little insight into how these changes came about and cannot test theories about the social processes underlying these changes, which are of particular interest for actor-oriented models \citep{snijders1996stochastic}.

The continuous-time evolution of our inferred network provides greater insight into potential social theories. In the right panels of Figure \ref{fig:MIT Snapshots} we plot snapshots of the inferred network at the same two survey dates. We observe similar behavior with student \#2, albeit with different individuals; we inferred three friendships in December and seven in March, corresponding to five formed friendships and one dissolved. Since the inferred network evolves in continuous-time, we can examine snapshots of the network at any time in between the survey dates to get a better sense of how the network evolves and see how the interactions influence the  formation/dissolution of inferred ties.

\begin{figure}[ht]
	\centering
	\includegraphics[scale=0.7]{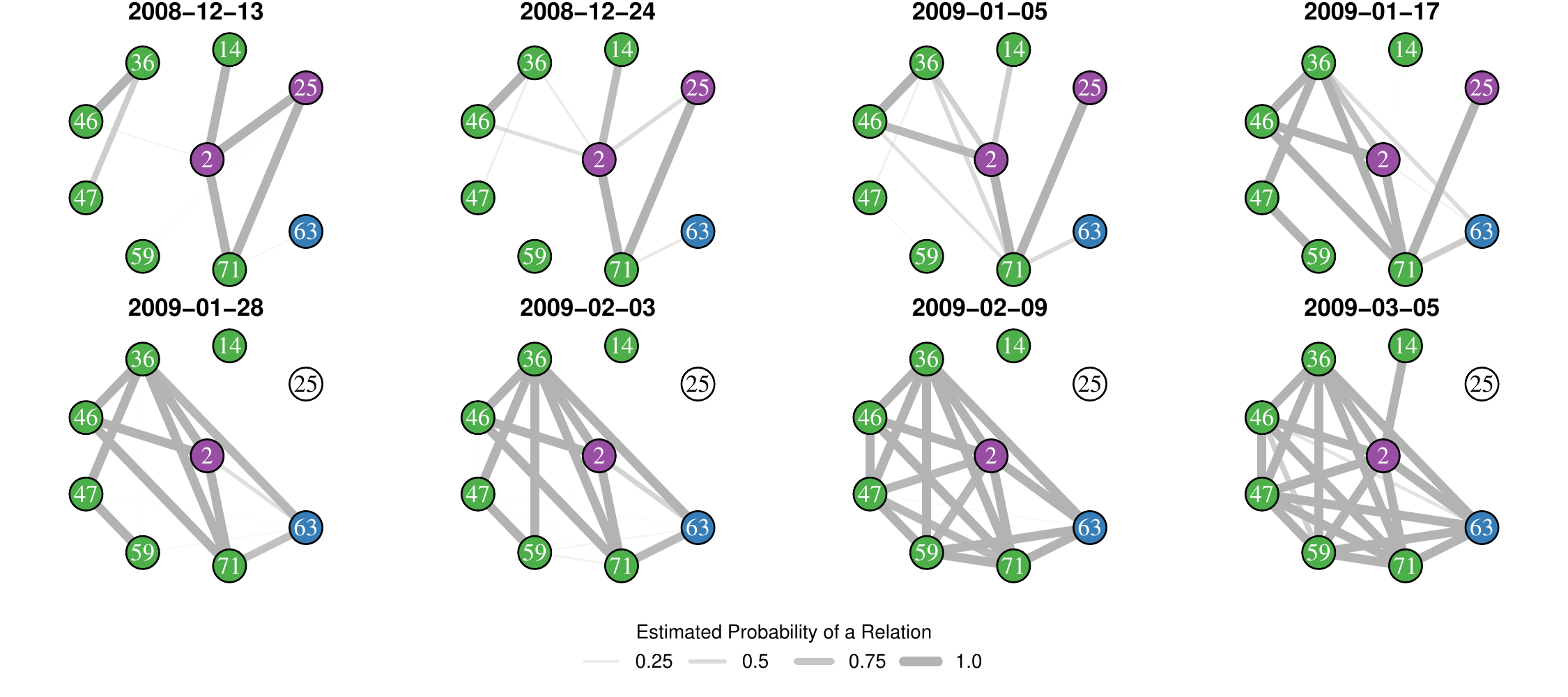} 
	\caption{Snapshots of the inferred network between December 13th, 2008 and March 5th, 2009 for a selected subset of students. Students are colored by the floor they lived on. Edge widths for each dyad denote the estimated mean probability of friendship at the given time.}
	\label{fig:MIT Egocentric Snapshots}
\end{figure}

In Figure \ref{fig:MIT Egocentric Snapshots}, we plot a subset of the inferred network for six additional dates between the survey dates. An animated series of all daily snapshots between the survey dates can be found at \url{https://git.io/viTtS}. These intermediate snapshots of the network provide evidence of triadic closure in this community, as friendships do not tend to form randomly but rather appear with higher propensity between actors who have mutual connections. We would have missed this behavior without the ability to examine the evolution of the network at finer resolutions; only once we were able to examine intermediate evolutions of the network were we able to find evidence of this phenomenon.

\section{Proximity Interactions among Barn Swallows} \label{Birds}

We now turn to the behavior of North American barn swallows (\textit{Hirundo rustica erythrogaster}). Unlike in the MIT student data, there is no survey data to guide our selection of the functions $w$, $m$, $a$, $s$, and $q$ underpinning the model.

\subsection{The Data} \label{Birds: Data}

Telemetry data were gathered for 17 barn swallows monitored from 06:00-09:00 and 17:00-20:00 (periods of high barn swallow activity) over three consecutive days during mating season \citep{levin2015performance,levin2016stress}. Barn swallows were outfitted with tracking devices which logged the presence of nearby swallows every 20 seconds. After data cleaning, detailed in Appendix~\ref{Swallow Appendix: Data}, the resulting dataset has 1009 interactions among its 136 dyads. Of the barn swallows studied, 10 were male and 7 were female. Various other physical characteristics of the birds are also available, including mass, tail streamer length, and plumage color.

\subsection{The Model} \label{Birds: Model}

We choose to omit various observable covariates known to be associated with barn swallow social behavior from our model.  For example, our model does not explicitly encode for factors that influence mate selection, such as the female preference to mate with males with darker plumage color  \citep{safran2004plumage,safran2005dynamic}.  Instead, our strategy is to develop a parsimonious model of the marginal network behavior without considering observables endogenous to network formation.  This strategy, which may be necessary for less well-studied communities, provides an overall characterization of barn swallow social behavior that is useful in its own right and, as we show in the results, can also be correlated with observable features after model fitting.

\begin{figure}[ht]
	\centering
	\includegraphics[scale=0.7]{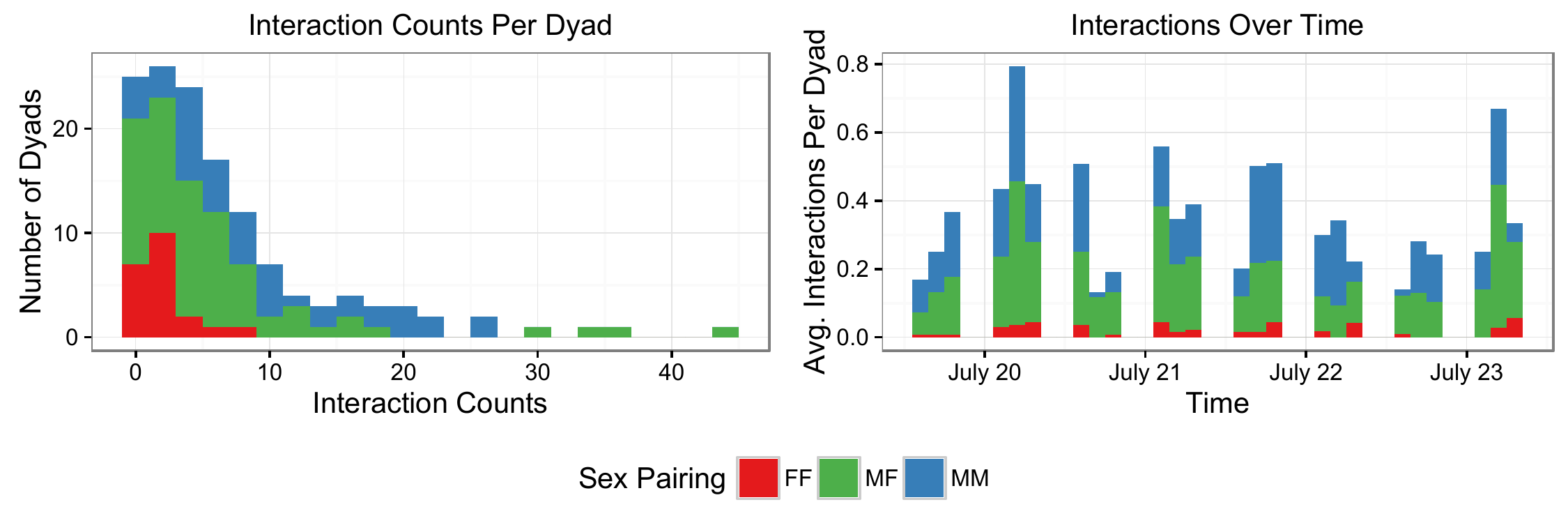}
	\caption{Descriptive statistics of the interaction data. Left: Histogram of the number of interactions for each dyad, color coded by the sex pairing of each dyad. Right:  Histogram of the number of interactions over time, aggregated across dyads. Interactions are also color coded by its dyads' sex pairing.}
	\label{fig:Swallow Descriptive}
\end{figure}

In Figure \ref{fig:Swallow Descriptive}, we see that both sex and time play important roles in understanding interaction behavior between swallows. The left panel shows that female birds interact very little with one another, yet interact quite often with males. In contrast, male birds regularly interact with each other. We take these differences in activity levels to represent behavioral discrepancies in the types of relationships formed by the various sex pairings rather than differences in the propensity for forming these relationships themselves. The right panel shows the rates of interaction during the periods of observation are non-constant. Differences in these rates may be due to factors outside of our interest, such  as the weather and other variable environmental stimuli, which are unlikely to impact network dynamics.

Under the assumption that swallows without social ties interact at a rate independent of their sex pairing, we choose to model the baseline activity rate for any pair at time $t$ as
\begin{equation}
	w\left(t\right)=k_{1}\mathbf{1}_{\text{7/19 PM}}
	+k_{2}\mathbf{1}_{\text{7/20 AM}}+...+k_{8}\mathbf{1}_{\text{7/23 AM}}.
	\label{eq:Barn Swallow Baseline}
\end{equation}
In this model, baseline activity rate is constant within each observational period but varies from period to period. This allows for flexibility in the temporal trend without imposing a strict parametric trend.

In contrast, we believe that barn swallows have different types of social relationships depending on the sexes of the birds involved and thus sex pairing will affect the activity levels of familiar birds. We embed this phenomenon by allowing for different multiplicative increases in interaction rates depending on sex pairing.  Omitting an additive effect $a$ to the baseline intensity, we assume that birds with social ties interact with the same temporal patterns as birds without ties. The intensity rate of the $\left(i,j\right)$th bird pair at time $t$ is then given by 
\begin{equation}
	\lambda_{ij}\left(t\right)=\big(1+
	\left(c_{FF}\mathbf{1}_{FF}+c_{MF}\mathbf{1}_{MF}+c_{MM}\mathbf{1}_{MM}
	\right)y_{ij}\left(t\right)\big)w\left(t\right),
	\label{eq:Barn Swallow Intensity}
\end{equation}
where $y_{ij}\left(t\right)$ is an indicator for whether or not 
$\left(i,j\right)$ are socially connected at time $t$. 

In accordance with our strategy to not explicitly encode observables into the formulation of the network, we model each dyad's connection $y_{ij}\left(t\right)$ with i.i.d. CTMCs with common sparsity $s$ and common transition parameter $q$. We complete the model by assigning weakly informative priors based on support matching to the parameters $\{k_1,...,k_8,c_{FF},c_{MF},c_{MM},s,q\}$. Detailed priors are given in Appendix~\ref{Swallow Appendix: Data}.

\subsection{Network Analysis} \label{Birds: Analysis}
We considered 9,000 samples (after burn-in) obtained from Stan (see Appendix~\ref{Swallow Appendix: Data} for computational details). A full animation of our inferred network, alongside the raw encounter network, is available at \url{https://git.io/viTtP}.

\begin{figure}[ht]
	\centering
	\includegraphics[scale=0.7]{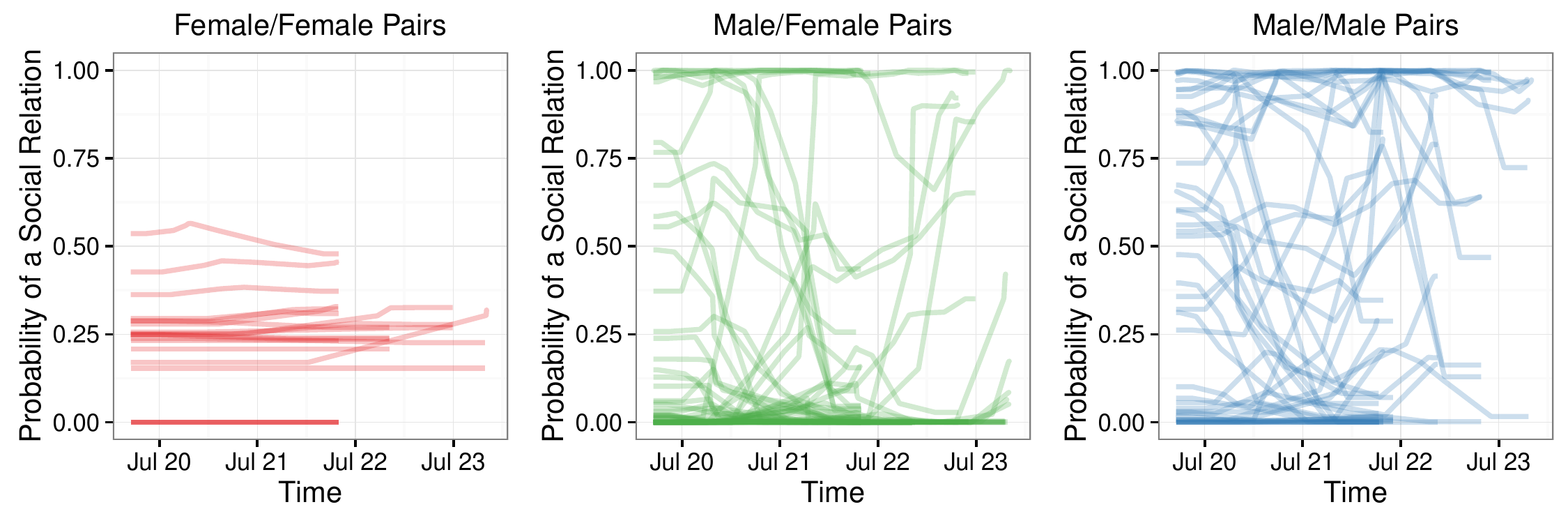}
	\caption{Estimated mean paths grouped by sex pairing.}
	\label{fig:Swallow Paths by Sex Pairing}
\end{figure}

In Figure \ref{fig:Swallow Paths by Sex Pairing}, we group the estimated mean network paths by sex pairing. In the left panel, we see the estimated social connections between female birds are generally static, without large changes in the probability of social ties over the course of the study. Unfortunately, some of the tracking devices on the female swallows became inactive early in the study, obscuring potential changes in the latter half of the study.

\begin{figure}[ht]
	\centering
	\includegraphics[scale=0.7]{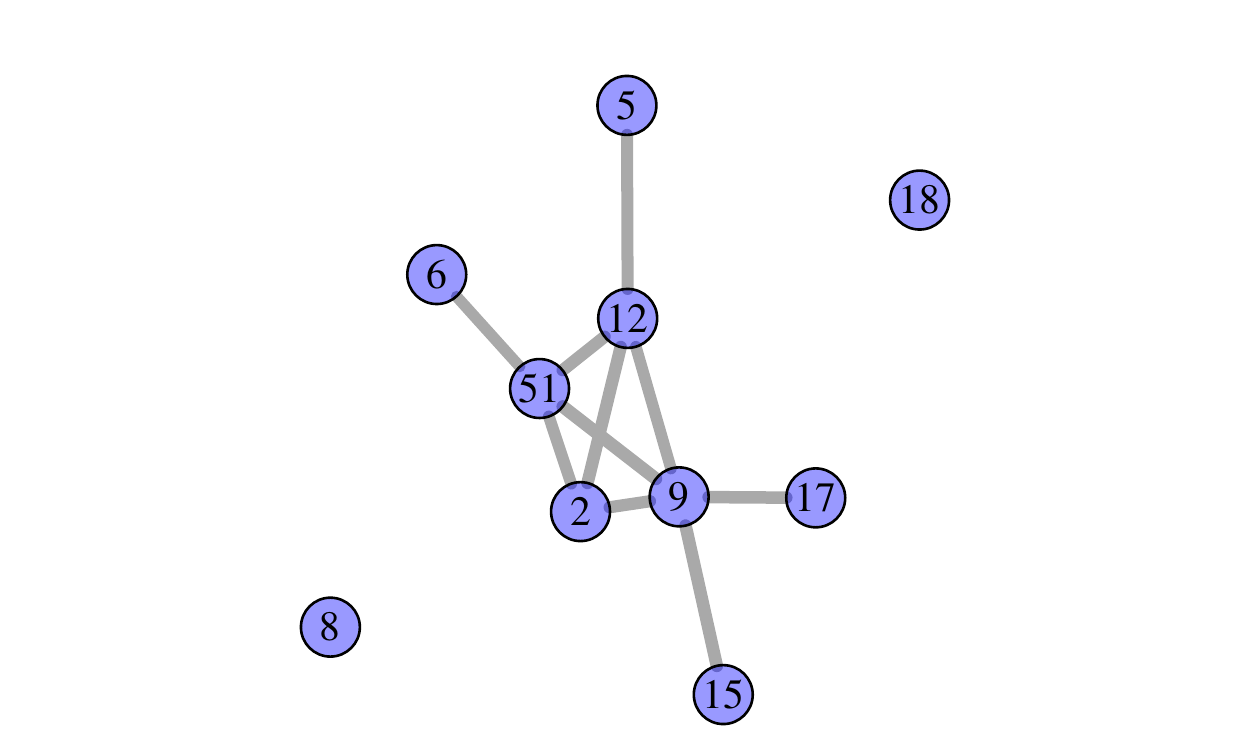}
	\caption{Network of male swallows. Nodes are labeled with each bird's tag identification number. Edges denote ``strong'' relationships, between birds who are predicted to be socially connected with probability 0.8 or higher on average over the study.}
	\label{fig:Swallow Male Subgraph}
\end{figure}

Displayed in the right panel of Figure \ref{fig:Swallow Paths by Sex Pairing}, pairs of male swallows exhibited a very diverse set of behaviors. In stark contrast to pairs of female swallows, many male pairs had dynamic associations in which their interactivity changed dramatically over the time of the study. The probability of a social tie fluctuated by at least 50\% over the three day period for slightly more than 20\% of possible male pairs. The differences in behavior may be attributable to the aggression of males during mating season, when they are particularly prone to engaging in territorial conflict with one another. At the same time, other pairs had relatively stable social ties or lack thereof. For example, we estimated 10 out of the possible 45 pairs to be socially connected with probability $0.8$ or higher when averaged over the course the study. These relations are plotted in Figure \ref{fig:Swallow Male Subgraph}.
We observe a clique of four male swallows, as well as four other males with fewer relations to the clique.  The stable behavior between these males may be evidence of extended territorial conflict or more complex social structure, such as unmated males ``attending" nests of other birds in the colony \citep{crook1987nonparental}.

Both strong stability and some dynamics characterize the behavior of male/female pairs, displayed in the middle panel of Figures \ref{fig:Swallow Paths by Sex Pairing}. We consistently classify about 75\% of the possible male/female pairings to have no relation. The remaining 25\% displayed a wide range of relational probabilities and dynamics. Among the latter were four disjoint pairs which averaged probability of relationships of 0.8 or higher. While barn swallows are sexually polygamous, they are socially monogamous and form pair-bonds for the breeding season; the behavior of these four pairs suggest that they may be such pairs.

\begin{figure}[ht]
	\centering
	\includegraphics[scale=0.7]{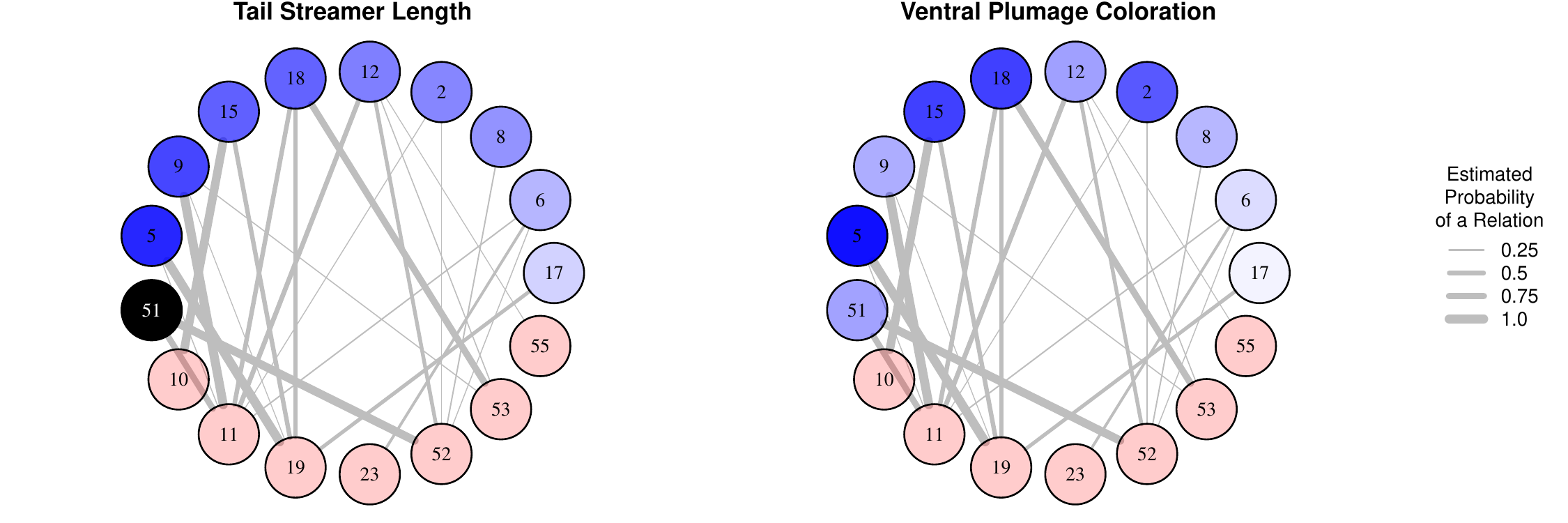}
	\caption{Male/female relationships among barn swallows. Red nodes denote females, while blue and black nodes denote males. Edges are weighted by average probability of a social relation, and estimated probabilities lower than 0.1 are omitted. In the left graph, males are shaded by tail streamer length, with darker color corresponding to longer tails (except for black, which corresponds to a missing value). In the right graph, males are shaded according to the color of their ventral plumage, with darker blue corresponding to darker coloration.}
	\label{fig:Swallow Sexual Selection}
\end{figure}

Previous studies of European barn swallows established that females prefer males with longer tail streamers \citep{moller1998sexual}, while in the United States it has been documented that females prefer males with darker ventral plumage \citep{safran2005dynamic}. Despite not encoding sexual selection into our model, we find evidence congruent with both theories for the studied North American barn swallows. Under the assumption that male/female relationships signal mating preferences, the males which were most successful in establishing bonds with females tended to have longer tail streamers and darker ventral plumage. Comparing the nine males for which tail streamer length was recorded, the three males with highest probability of relations with females had the three longest tails (among the nine). This finding, illustrated in Figure \ref{fig:Swallow Sexual Selection}, is quite robust to the choice of cutoff dividing males by tail streamer length. A similar relationship held to a lesser extent for ventral coloring, which was strongly correlated ($r$=0.79) with tail streamer length for the studied males.

\section{Discussion} \label{Discussion}

The proliferation of continuous-time relational event data presents both new opportunities and challenges for social network research. Event data represent observed behavior, which fundamentally differs from reported behavior, the traditional domain of survey data \citep{killworth1976informant,eagle2009inferring}. Many of the network analysis tools developed for survey data are not appropriate for relational event data \citep{howison2011validity}, and new methods must be proposed. Simultaneously, a better understanding of the data reliability issues surrounding the data collection process is required.  Most importantly, these data expose numerous ethical issues related to how the data should be used and how to ensure privacy of individuals. 

We have presented a hierarchical network model in which relational events are modeled with Poisson processes governed by dyadic covariates and a dynamic latent network. Our model acknowledges the intrinsic nature of relational event data as continuous-time interactions only indirectly indicative of underlying dynamic network relations. The concept of differentiating between spurious patterns of behavior and those indicative of a more meaningful connection is fundamental to many applications. In information security, botnets are networks of devices infected by software which execute automated scripts, often for malicious purposes. We can monitor and investigate the spread of a botnet by looking for deviations from typical communication patterns between devices \citep{pao2015statistical}. In over-the-counter markets, which trade financial instruments without the use of centralized exchanges, parties directly deal with one another and are subject to less regulatory oversight. As with stock exchanges \citep{landers1992complexity}, examining the behavior of these markets at a micro-level, specifically by modeling individual trades between the various parties, can lend insight into the relational structure of the parties in these markets and the potential consequences of this structure, such as the role of the interconnectivity of financial institutions in exacerbating the 2008 financial crisis.

Depending on the intended application, further developments of the model may allow for more precise estimates of the latent network or more fully capturing dependence between events. For example, in organizational email data, emails can have multiple recipients, deviating from our dyadic framework, and each email is associated with content that informs the nature of the interaction. Differentiating between these types of interactions through the use of multiple connected point processes may yield more accurate estimates of our network of interest. Alternatively, we may want our model to discriminate between the various types of relationships between workers through the use of multiple latent networks. Economic phenomena, such as crisis management and information diffusion, may occur through different channels of interest. As another example, when modeling botnets, mutually-exciting Hawkes processes may be more suitable than independent Poisson processes for modeling the spread of malware, since malware can only be transmitted from devices that are already themselves infected. The structure of relational event data, its relationships to an underlying network, and the meaning of the network will vary considerably across applications.  Hence, domain knowledge should be used whenever possible to customize the model to target capturing relevant and interesting structure.\\

\appendix
\section{Appendix: MIT Social Evolution Data} \label{MIT Appendix}

\subsection{Data Considerations} \label{MIT Appendix: Data}

Participants in the study were given smartphones outfitted with an application that used Bluetooth to scan for other nearby (within 10 meters) smartphones approximately every six minutes. However, the application only logged the presence of other devices set as ``discoverable.'' Due to this and other technical issues, many of the reported encounters were non-reciprocal and it is possible many encounters went unreported. In addition, a fraction of the logged encounters occurred between phones on different floors due to vertical proximity. These encounters are not of interest, and are identifiable as being estimated by the application to have low probability of being on the same floor using WLAN location data. We chose to drop all encounters with probability 0.2 or lower. In addition, we only considered interactions that occur during the Fall (September 3rd to December 13th) and Spring sessions (January 5th to March 20th and March 30th to May 15th), notably excluding winter and spring breaks, which were extended periods when school is not in session. Please refer to \cite{madan2012sensing} for further details about the data.

We converted the proximity logs into undirected (reciprocal) interaction data by combining the directed proximity logs for pairs of individuals, taking encounters logged in either phone. Logs that were time-stamped within 30 minutes of one another were taken to correspond to the same interaction.

Defining a dyad to have reported as friends in a survey if either marks the other as a `close friend,' we observed significant dissonance between the survey data and interaction data on a dyadic level. Over twenty percent of dyads who reported friendship in all five surveys did not have any recorded interactions, compared to just under fifty percent for dyads who never reported friendship. The weak relation between the sources of data was robust to different choices of defining friendship in the survey data and raises validity concerns about both sets of data.

\begin{figure}[ht]
	\centering
	\includegraphics[scale=0.7]{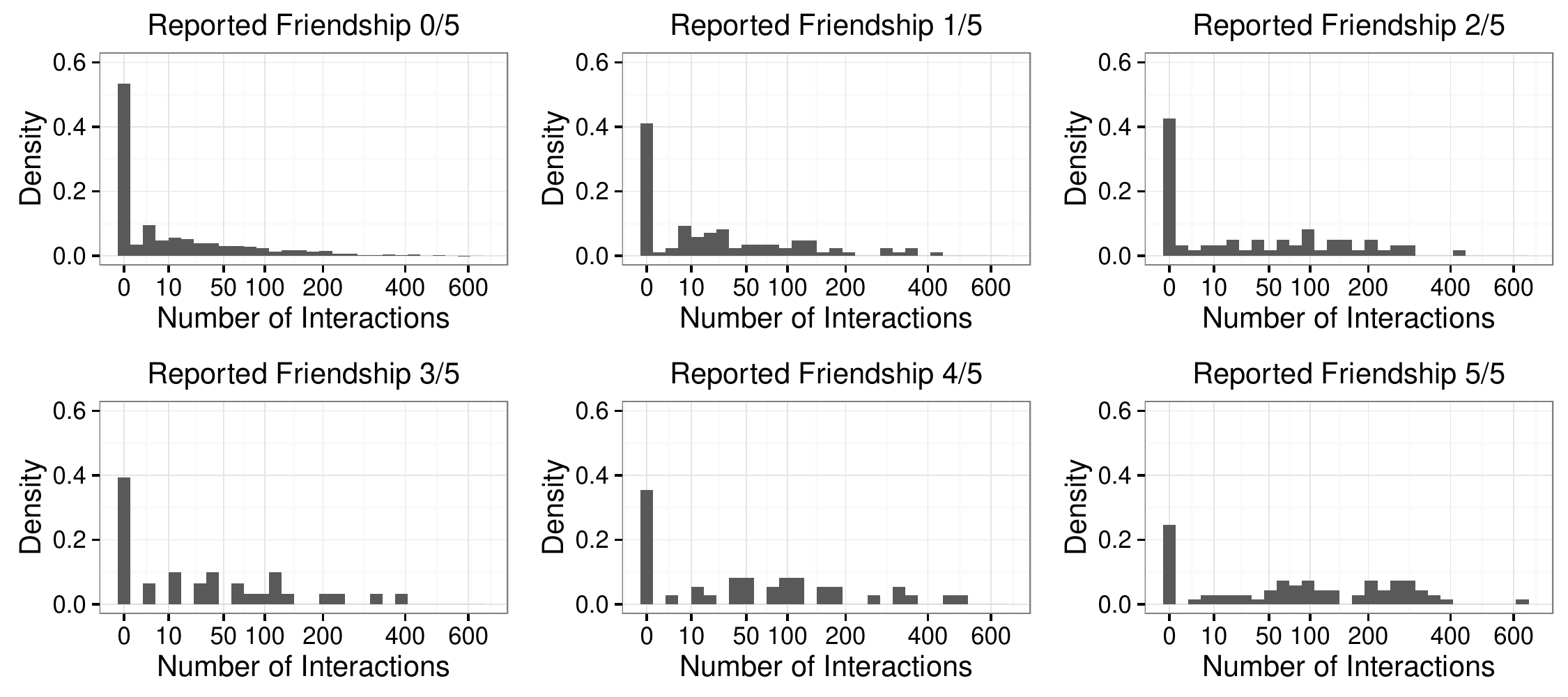}
	\caption{Distributions of the total number of recorded interactions per dyad, where the data is partitioned by the fraction of times each dyad reported friendship in the five surveys. We consider a dyad to have reported as friends in a survey if either individuals marks the other as a `close friend.'}
	\label{fig:MIT Ints vs Surveys}
\end{figure}

\subsection{Priors and Model Fitting} \label{MIT Appendix: Model}

We assign the following independent priors to the parameters of the model specified by \eqref{eq:MIT Baseline}-\eqref{eq:MIT Friendship Sparsity}): (where applicable, units are in hours)
\begin{align*}
	\{c_{0,\text{t2}},c_{0,\text{t3}}\} & \sim \text{Exp}\left(1\right) & 
	c_{1} & \sim \text{Exp}\left(1\right)\\
	\{c_{2,1},c_{2,2}\} & \sim \text{Exp}\left(100\right) &
	c_{3} & \sim \text{Exp}\left(1\right)\\
	k_{0} & \sim \text{Exp}\left(1\right) &
	\{k_{i,1}:i\in\{1,2,3\}\} & \sim \text{Exp}\left(1\right)\\
	\{k_{i,3}:i\in\{1,2,3\}\} & \sim \text{Uniform}\left(0,2\pi\right) &
	s_{0} & \sim \text{Beta}\left(1,49\right)\\
	s_{1} & \sim \text{Exp}\left(1\right) &
	s_{2} & \sim \text{Exp}\left(1\right)\\
	q & \sim \text{Exp}\left(1e10\right)	
\end{align*}

\begin{figure}[hp]
	\centering
	\includegraphics[scale=0.7]{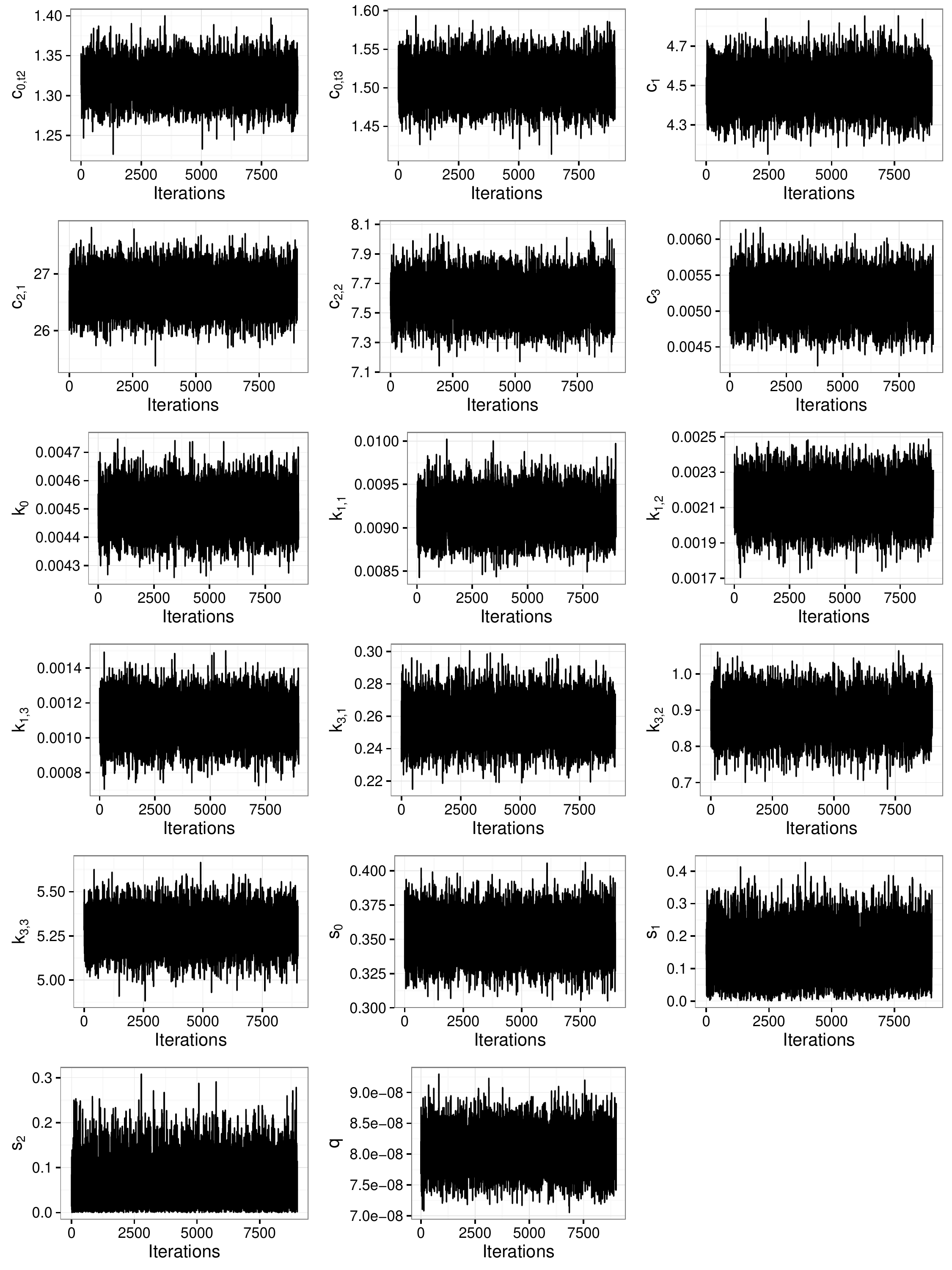}
	\caption{Trace plots for the MIT model parameters. Values obtained during the burn-in period are discarded.}
	\label{fig:MIT Trace}
\end{figure}

We used Stan to generate 10,000 samples using a single MCMC chain, initializing each of the parameters at its prior mean and discarding the first 1,000 as burn-in. This sampling procedure took approximately 24 hours on a standard laptop. We assessed convergence by visually examining the trace plots shown in Figure \ref{fig:MIT Trace}.

\section{Appendix: Barn Swallow Data}  \label{Swallow Appendix}

\subsection{Data Considerations} \label{Swallow Appendix: Data}

Every 20 seconds, the proximity loggers attached to each barn swallow logged the presence of other barn swallows within 5 meters. This threshold was chosen by the researchers since it represented the average distance between nests at the observation site. The devices split logged encounters between birds that lasted more than 5 minutes into 5 minute intervals. Unfortunately, the proximity loggers were imperfect; sometimes encounters recorded by one bird were not recorded by the other and the batteries of some devices died (starting on July 22nd) before the end of the observation period. See \cite{levin2015performance} for a more comprehensive description of the data.

We accounted for asymmetries and extended encounters in the encounter log. To ensure encounters that last longer than 5 minutes were treated as a single interaction, encounters separated by thirty seconds or less were merged. To compensate for the imperfect behavior of the tracking devices, we combined the logged encounter history for every pair of birds, taking encounters logged in either birds' tracker as an interaction and merging overlapping encounters. Lastly, to help avoid issues related to battery-life, we only modelled each dyad over the interval when the trackers of both birds recorded activity.

\subsection{Priors and Model Fitting} \label{Swallow Appendix: Model}

We assign the following priors to the parameters of our model: (units for $k_{i}$ and $q$ are provided in hours)
\begin{align*}
	\{c_{FF},c_{MF},c_{MM}\} & \sim \text{Exp}\left(2\right) &
	\{k_{i}: i\in\{1,...,8\}\} & \sim \text{Exp}\left(1\right) \\
	s & \sim \text{Beta}\left(1,9\right) &
	q & \sim \text{Exp}\left(1000\right)	
\end{align*}

\begin{figure}[hp]
	\centering
	\includegraphics[scale=0.7]{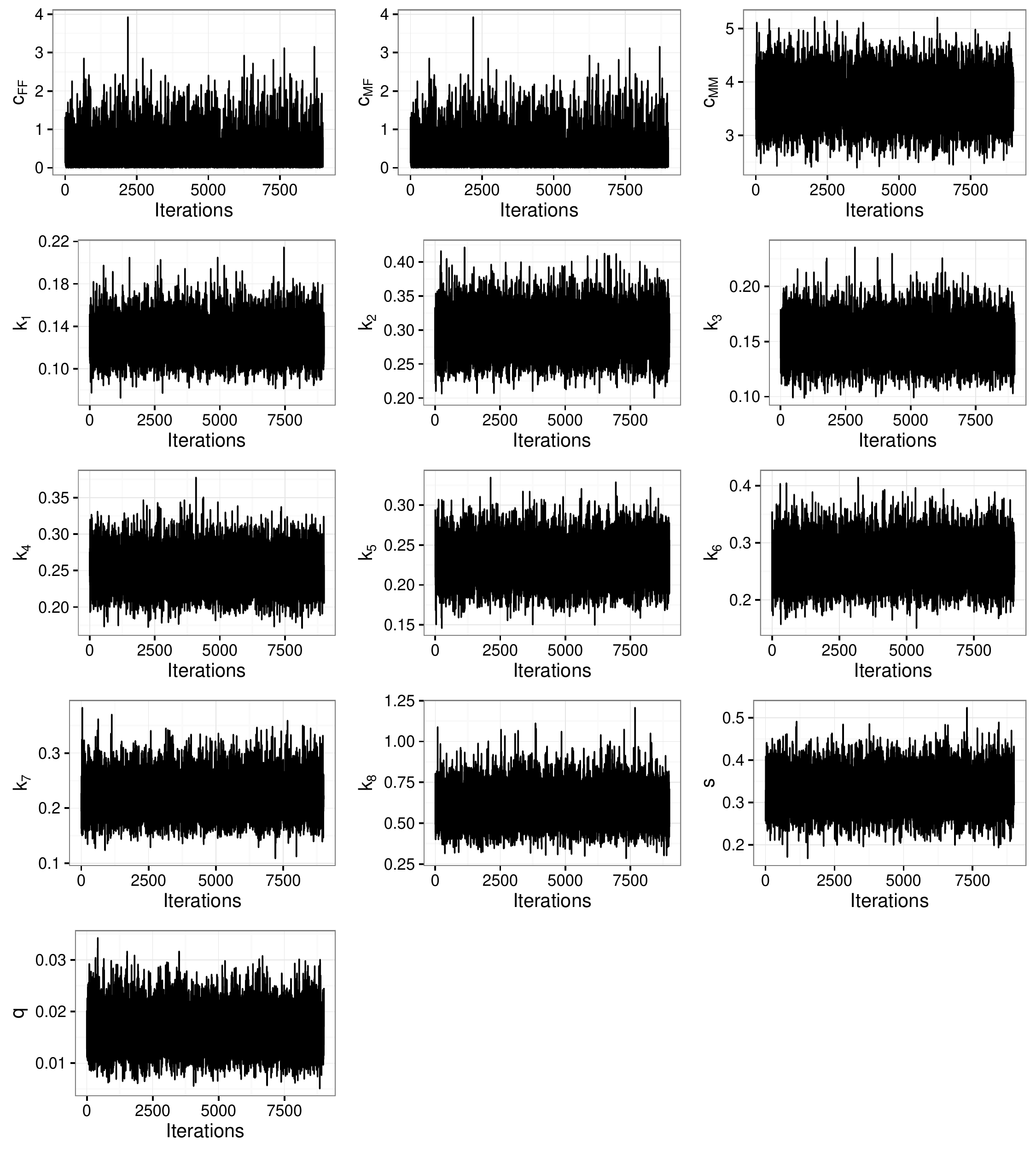}
	\caption{Trace plots for barn swallow model parameters. Values obtained during the burn-in period are discarded.}
	\label{fig:Swallow Trace}
\end{figure}

Using Stan, we ran a single MCMC chain for 10,000 iterations and discarded the first 1,000 as burn-in. The parameters were initialized at the means of their respective prior distributions, and the entire sampling procedure took approximately 6 minutes. As with the previous application, we assessed convergence by examining trace plots, provided in Figure \ref{fig:Swallow Trace}.

\section*{Acknowledgements}
The authors would like to thank Iris Levin and Henry Scharf for helpful comments and suggestions on an earlier version of the manuscript, and Iris Levin and Rebecca Safran for providing the barn swallow interaction data.  This work was partially supported by the National Science Foundation under Grant Number SES-1461495 to Fosdick and Grant Number SES-1559778 to McCormick. Additional support from grant number K01 HD078452 from the National Institute of Child Health and Human Development (NICHD).  This material is based upon work supported by, or in part by, the U. S. Army Research
Laboratory and the U. S. Army Research Office under grant number W911NF-12-1-0379.

\bibliographystyle{asmbi}
\bibliography{references2}

\end{document}